\documentclass[journal=nalefd,manuscript=letter]{achemso}

\usepackage{graphicx}% Include figure files
\usepackage{dcolumn}% Align table columns on decimal point
\usepackage{bm}% bold math
\usepackage{multirow} %For multirows in Tables
\usepackage[separate-uncertainty = true]{siunitx}
\usepackage[Symbol]{upgreek} %For upright pi
\usepackage{multirow}

\newcommand{\RN}[1]{\uppercase\expandafter{\romannumeral#1}}

\author{Hendrik K\"ahler}
\affiliation{Institute of Sensor and Actuator Systems, TU Wien, Gusshausstrasse 27--29, 1040 Vienna, Austria.}
\author{Holger Arthaber}
\affiliation{Institute of Electrodynamics, Microwave and Circuit Engineering, TU Wien, Gusshausstrasse 25, 1040 Vienna, Austria.}
\author{Robert Winkler}
\affiliation{Christian Doppler Laboratory for Direct-Write Fabrication of 3D Nanoprobes (DEFINE), Institute of Electron Microscopy and Nanoanalysis, Graz University of Technology, Steyrergasse 17, 8010 Graz, Austria.}
\author{Robert G. West}
\affiliation{Institute of Sensor and Actuator Systems, TU Wien, Gusshausstrasse 27--29, 1040 Vienna, Austria.}
\author{Ioan Ignat}
\affiliation{Institute of Sensor and Actuator Systems, TU Wien, Gusshausstrasse 27--29, 1040 Vienna, Austria.}
\author{Harald Plank}
\affiliation{Christian Doppler Laboratory for Direct-Write Fabrication of 3D Nanoprobes (DEFINE), Institute of Electron Microscopy and Nanoanalysis, Graz University of Technology, Steyrergasse 17, 8010 Graz, Austria.}
\affiliation{Institute of Electron Microscopy and Nanoanalysis, Graz University of Technology, Steyrergasse 17, 8010 Graz, Austria.}
\affiliation{Graz Centre for Electron Microscopy, Steyrergasse 17, 8010 Graz, Austria.}
\author{Silvan Schmid}
\email{silvan.schmid@tuwien.ac.at}
\affiliation{Institute of Sensor and Actuator Systems, TU Wien, Gusshausstrasse 27--29, 1040 Vienna, Austria.}

\title{Transduction of single nanomechanical pillar resonators by scattering of surface acoustic waves}

\begin{document}

\begin{abstract}
%In the last decades, the dimensions of microelectromechanical systems (MEMS) have been pushed to the nanometer range to improve the systems' sensitivity for sensing applications.
One of the challenges of nanoelectromechanical systems (NEMS) is the effective transduction of the tiny resonators. Vertical structures, such as nanomechanical pillar resonators, which are exploited in a wide range of fields, such as optomechanics, acoustic metamaterials, and nanomechanical sensing, are particularly challenging to transduce. Existing electromechanical transduction methods are ill-suited as they complicate the pillars' fabrication process, put constraints on the pillars' material, and do not enable a transduction of freestanding pillars. Here, we present an electromechanical transduction method for single nanomechanical pillar resonators based on surface acoustic waves (SAWs). We demonstrate the transduction of freestanding nanomechanical platinum-carbon pillars in the first-order bending and compression mode. Since the principle of the transduction method is based on resonant scattering of a SAW by a nanomechanical resonator, our transduction method is independent of the pillar's material and not limited to pillar-shaped geometries. It represents a general method to transduce vertical mechanical resonators with nanoscale lateral dimensions.

%, enabling applications in mass and force sensing.
\end{abstract}

%\maketitle

\section{Introduction}
Micro- and nanomechanical pillar resonators are extremely versatile due to their vertical structure and capability to be arranged in dense arrays.  Pillar resonators allow for the mass detection of nanoparticles\cite{Wasisto2013,Bonhomme2019}, the sensing of forces\cite{Rossi2017,DeLepinay2017,Nichol2012}, the strong confinement of photons and phonons\cite{Anguiano2017,Asano2020}, and the manipulation of quantum dots\cite{Yeo2014,Kettler2021,Wigger2018} and surface acoustic waves (SAWs)\cite{Jin2021,Raguin2019,Achaoui2011,Liu2019,Oudich2010,Pennec2009}, which are both exploited for quantum information processing\cite{Aharonovich2016,Bienfait2019}. However, many of the common electrical transduction methods used for horizontally designed nanoelectromechancical systems (NEMS) are not convenient for vertical pillar resonators, such as piezoresisitive\cite{Tortonese1993,Li2007}, piezoelectric\cite{OConnell2010,Karabalin2009}, electrothermal\cite{Bargatin2007}, and magnetomotive transduction\cite{Cleland1996,Feng2008}. These methods rely on electrodes directly placed on top of the mechanical resonator, which cannot be done for pillars with standard lithographic fabrication techniques. That limits the feasible electrical transduction methods to capacitive transduction\cite{Sazonova2004,Bunch2007,Lassagne2008,Truitt2007}  and transduction by dielectric forces\cite{Unterreithmeier2009,Schmid2006}. Both were successfully used for pillar resonators\cite{Montague2012,Toffoli2013}, but electrodes have to be placed close to the mechanical resonator for both transduction methods. This comes with two disadvantages. First, the electrodes have approximately the same height as the pillars \cite{Toffoli2013,Montague2012}, which considerably complicates the fabrication process. Second, the pillars are not freestanding, which is unfavorable for sensing applications, such as force sensing and particle mass detection. Apart from pure electrical transduction methods, optical methods\cite{Raguin2019,Rossi2017,Molina2020} and scanning electron microscopy (SEM)\cite{Wasisto2013,Doster2019} have been used to detect the motion of single pillars. These approaches have the advantage that the pillars are freestanding, but they are difficult to integrate. 

Here, we demonstrate a transduction method for single pillar resonators, which combines the advantages of electrical and optical transduction methods by using SAWs. The SAWs are launched and detected by interdigital transducers distanced hundreds of micrometers away from the pillar resonator. This enables a transduction of freestanding pillars and reduces constraints on the pillars' fabrication process. 

\section*{SAW transduction scheme}
\begin{figure}
  \begin{center}    \includegraphics{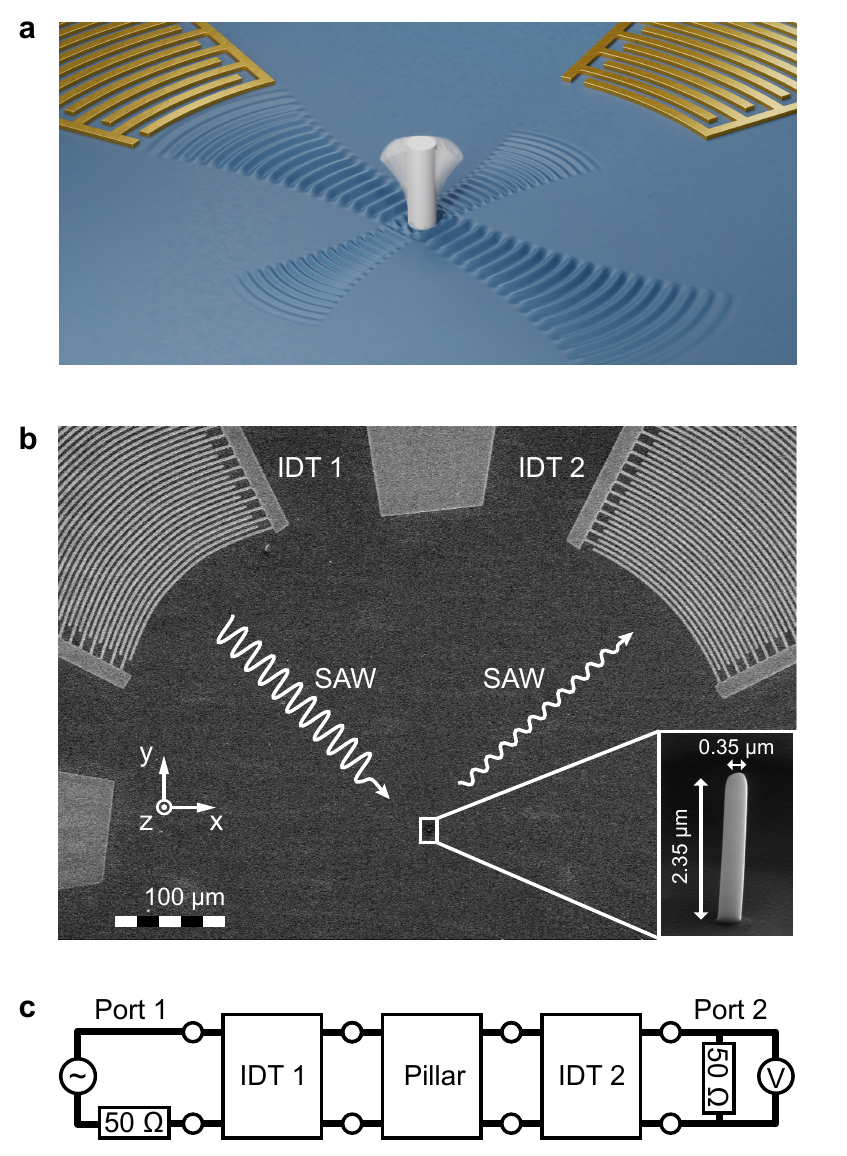}
    \caption{Surface acoustic wave (SAW) transduction scheme. (a) Illustration of the SAW transduction. The key components of the device are a piezoelectric substrate, two interdigital transducers (IDTs) and a pillar resonator, which are coloured in blue, gold, and grey, respectively. One IDT emits a SAW to drive the pillar resonator in the center. The other IDT detects the motion of the pillar by measuring the SAW scattered by the pillar resonator at resonance. The pillar vibrates in its first bending mode. (b) Scanning electron microscope image of a device used in this study. The white, wavy lines represent the SAWs, which are launched and detected by the interdigital transducers (IDTs). (c) Equivalent circuit model. The device is represented by a cascade of two-port networks.}
    \label{ElecSetup}
  \end{center}    
\end{figure}

A schematic and a SEM image of a device used in this study are shown in Fig.~\ref{ElecSetup}a, b. The device consists of two perpendicularly oriented interdigital transducers (IDTs) and a single pillar resonator. As a substrate, we used piezoelectric lithium niobate (LiNbO$_3$) with a $128^\circ$~Y-cut orientation. The IDTs and the pillars were fabricated by photolithography and Focused Electron Beam Induced Deposition (FEBID) out of a platinum metal-organic precursor \cite{Winkler2019}, respectively. The IDTs convert an electrical input signal to a SAW and vice versa. They are optimized for the generation and detection of Rayleigh-type SAWs. One of the IDTs launches a SAW to drive the pillar resonator and the other IDT measures the SAW created by the pillar's motion, as illustrated in Fig.~\ref{ElecSetup}a. To maximize the signal strength, the electrodes of the IDTs are designed to follow the shape of the wave surface of the SAW\cite{Laude2008,Benchabane2017,ORorke2020} that is emitted by the pillar. Each IDT covers an angle of $35^\circ$.

In the following, we discuss the results of three devices with pillars of different dimensions. We refer to the them as the thin, the midsize, and the wide pillar. The dimensions of all pillars are given in Table~\ref{Table} as well as the central frequency and bandwidth of the corresponding IDTs. The bandwidth of the IDTs is defined by a reduction in output power by \SI{-3}{\decibel}.

\setlength{\tabcolsep}{4pt}
\renewcommand{\arraystretch}{1.25}
\begin{table}[t]
\caption{Key parameters of the devices used in this study. The parameters $d$ and $h$ are the pillars' diameter and height, respectively, and the parameters $f_\text{c}$ and $BW$ are the central frequency and bandwidth of the used IDTs, respectively.}
\centering
\begin{tabular}{ c c c | c c }
 \hline
 \hline
 \multicolumn{3}{c|}{Pillar} & \multicolumn{2}{c}{IDT} \\ 
Name & $d$~(\SI{}{\micro\meter}) & $h$~(\SI{}{\micro\meter}) & $f_\text{c}$~(\SI{}{\mega\hertz}) & $BW$~(\SI{}{\mega\hertz}) \\
 \hline
 Thin & 0.35 & $2.35\pm0.05$ & 280 & 110 \\
 Midsize & 0.70 & $2.40\pm0.05$ & 280 & 110\\
 Wide & 2.20 & $1.7\pm0.1$ & 178 & 70 \\
 \hline
 \hline
\end{tabular}
\label{Table}
\end{table}

\section*{Equivalent circuit model}
In the field of microwave engineering, complex circuits are often modelled by two-port networks and described by so called scattering parameters, such as SAW devices\cite{Morgan2007}. Scattering parameters represent ratios of outgoing to incoming normalized power waves and hence are well-suited to describe the scattering of a SAW by a pillar in an equivalent circuit model. Such a model of the SAW transduction scheme is shown in Fig.~\ref{ElecSetup}c. The two IDTs and the single pillar are each modelled by a two-port network defined by scattering parameters. The travelling of the incoming and scattered SAWs are included in the respective IDT networks. In a cascade of two-port networks power waves can travel back and forward between the single networks, which complicates the calculation of the overall transmission scattering parameter $S_\text{21}$. However, it can be assumed that backscattering between the networks is minimal due to low reflection of SAWs at the IDTs and the pillar. The IDTs are chirped and the diameters of the pillars are less than a tenth of the SAW's wavelength at the pillars' resonances. In this case, the overall transmission scattering parameter $S_\text{21}$ is given by  
\begin{equation}
S_\text{21}(f) = S_\text{IDT,1}(f) \, S_\text{P}(f) \, S_\text{IDT,2}(f) \; ,
\label{S21}
\end{equation}
where $f$ is the frequency of the applied input signal, and $S_\text{IDT,i}$ and $S_\text{P}$ are the transmission scattering parameters of the IDTs and the pillar, respectively. We make a distinction between the two IDTs, since they are placed along different crystalline axis of the LiNbO$_\text{3}$ substrate which results in a different electromechanical coupling of the IDTs to the substrate\cite{Zhang2020}. However, the IDTs are designed equivalently. If we assume that the electromechanical coupling of an IDT to the substrate only determines the amplitude of the transmission of an IDT, since the frequency characteristic of an IDT is mainly given by the distances of its electrodes\cite{Morgan2007}, the transmission scattering parameters of the two IDTs are proportional to each other and $S_\text{21}$ simplifies to
\begin{equation}
S_\text{21}(f) = C \, S_\text{IDT}^2(f) \, S_\text{P}(f) \; ,
\label{S21simple}
\end{equation}
where $C$ is a constant.

The scattering of the SAW by the pillar resembles the scattering of light by resonating objects with dimensions much smaller than the optical wavelength, such as molecules or metallic nanoparticles. The scattering cross section $\sigma_\text{scat}$ of such scattering processes is given by\cite{Hamam2007,Lee2016}
\begin{equation}
\sigma_\text{scat}(f) \propto (\frac{1}{Q_\text{rad}})^2  \, \frac{f^4}{(f_0^2-f^2)^2+ f^2 \, (\frac{f_0}{Q_0})^2} \; , 
\label{scatcross}
\end{equation}
where $f$ is the frequency of the light, and $f_0$, 
$Q_0$ and $Q_\text{rad}$ are the eigenfrequency, the total and radiation quality factor of the resonating object, respectively. Scattering cross sections describe the ratio of the scattered power to the intensity of the incident wave. In contrast, scattering parameters are defined by the square root of the incoming to outgoing power. Having this in mind, $S_\text{P}$ is given from Eq.~(\ref{scatcross}) as follows
\begin{equation}
S_\text{P}(f) = S_\text{eff} \, \frac{f^2}{f_0^2-f^2+\text{i} \, f \, \frac{f_0}{Q_0}} \; , 
\label{Sp}
\end{equation}
where $S_\text{eff}$ is an effective scattering parameter. $S_\text{eff}$ is proportional to $1/Q_\text{rad}$ and is quite likely also a function of frequency. The SAW scattered by the pillar is focused by the IDT and it can be expected that the beam width of the SAW changes with frequency. Additionally, the penetration depth of the SAW into the substrate is a function of frequency\cite{Morgan2007,Classen1995}. However, we assume $S_\text{eff}$ to be constant in the following, since we are only interested in the narrow frequency regions around the resonances of the pillars. 

\begin{figure*}[t]
  \begin{center}\includegraphics{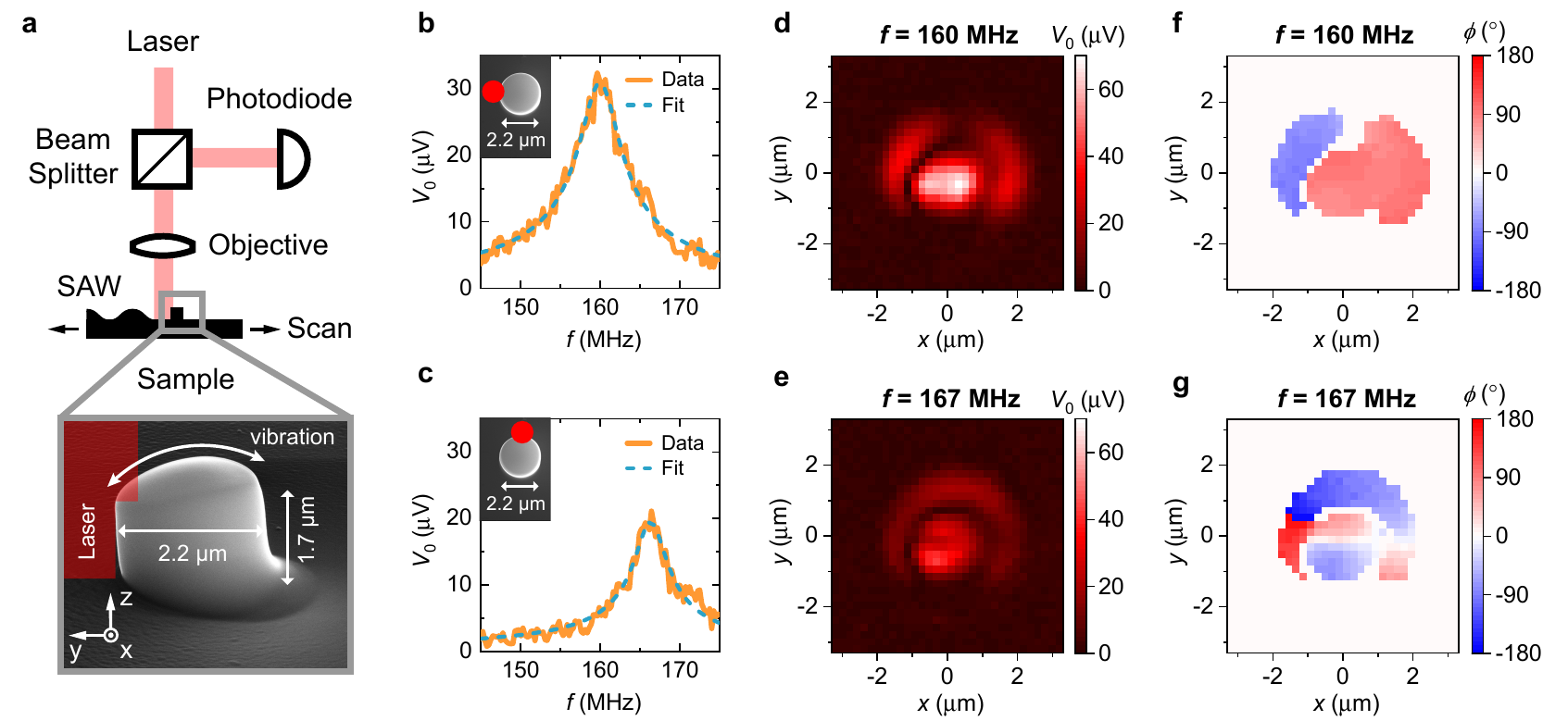}%[width=1\textwidth]
    \caption{Optical detection of the motion of a pillar resonator. (a) Schematic of the optical setup and a scanning electron microscope image of the investigated pillar. The given height of the pillar is its average height. The pillar was driven by a surface acoustic wave (SAW). (b), (c) Frequency response of the pillar for two different laser positions. We measured the amplitude of the photodiode's output signal $V_0$ at the applied SAW's frequency. The insets show the pillar in top view. We fitted the frequency response of a driven, weakly-damped harmonic oscillator to the data. (d), (e) Amplitude and (f), (g) phase of the optical signal for fixed frequencies as a function of the laser position with a resolution of \SI{200}{\nano\meter}. For clarity, we only show the phase of the optical signal at the laser positions, where the optical signal is above the noise level. The pillar is located around the center of the maps.}
    \label{OptSetup}
  \end{center}    
\end{figure*}

\section*{Optical characterization}
In addition to the transduction only by SAWs, we investigated optically the motion of the wide pillar induced by SAWs. A schematic of the optical detection setup is shown in Fig.~\ref{OptSetup}a. The optical signal is generated by scattering of the incident and reflected light by the lateral motion of the pillar, as demonstrated by Molina et. al.\cite{Molina2020}. The frequency responses of the wide pillar for two different laser positions on the edge of the pillar are shown in Fig.~\ref{OptSetup}b,c. It can be seen that the incident SAW excites two eigenmodes of the pillar: one at \SI{160}{\mega\hertz} with a quality factor of $Q=32$ and the other at \SI{167}{\mega\hertz} with a quality factor of $Q=41$. 

To determine the type of these two modes, we measured the amplitude and phase of the optical signal for the two frequencies as a function of the laser position. The results are given in Fig.~\ref{OptSetup}d-g and correspond with the results of Molina et. al.\cite{Molina2020} (see Supporting Data section S1). Two orthogonal bending modes are clearly visible: one mode vibrating along the x-direction and the other along the y-direction. Both modes show a phase difference of $180^\circ$ between the opposite sides of the pillar, which is typical for bending modes. The relatively large frequency difference between the two orthogonal bending modes originates from the geometrical asymmetry of the pillar, as can be seen in Fig.~\ref{OptSetup}a. The pillar shows a ramp on one side of its base. The ramp is a result of a drift of the electron beam at the start of the pillar's writing process due to charging effects and causes a reduction of the signal amplitude in negative y-direction, as can be seen in Fig.~\ref{OptSetup}e.

\section*{FEM simulations}
We compared the optical results to finite element method (FEM) simulations by simulating the eigenmodes of the wide pillar. The material properties of the pillar were based on previous studies\cite{Arnold2018,Utke2020,Benchabane2017}. We set the Young's Modulus, mass density, and Poisson's ratio to $E = (25\pm15)~\text{GPa}$, $\rho = 4000~\text{kg/m}^3$ and $\nu=0.38$, respectively. The relatively large range of the Young's modulus includes the possibility that the pillar experienced e-beam curing during its fabrication process\cite{Plank2013}. E-beam curing describes the chemical modification of FEBID-fabricated, platinum-carbon pillars that are exposed to high doses of electrons, resulting in a significant increase of the Young's modulus\cite{Arnold2018}. Based on the study of Arnold et al.\cite{Arnold2018}, we expect an e-beam curing of our pillars for two reasons. First, we used larger e-beam currents of \SI{91}{\pico\ampere} for fabrication of the pillars compared to Arnold et al. Second, our pillars are relatively wide in comparison to the electron interaction volume\cite{Winkler2021}, so that electrons scattered horizontally in the pillar contribute to curing as well. 

\begin{figure}[t]
  \begin{center}   \includegraphics{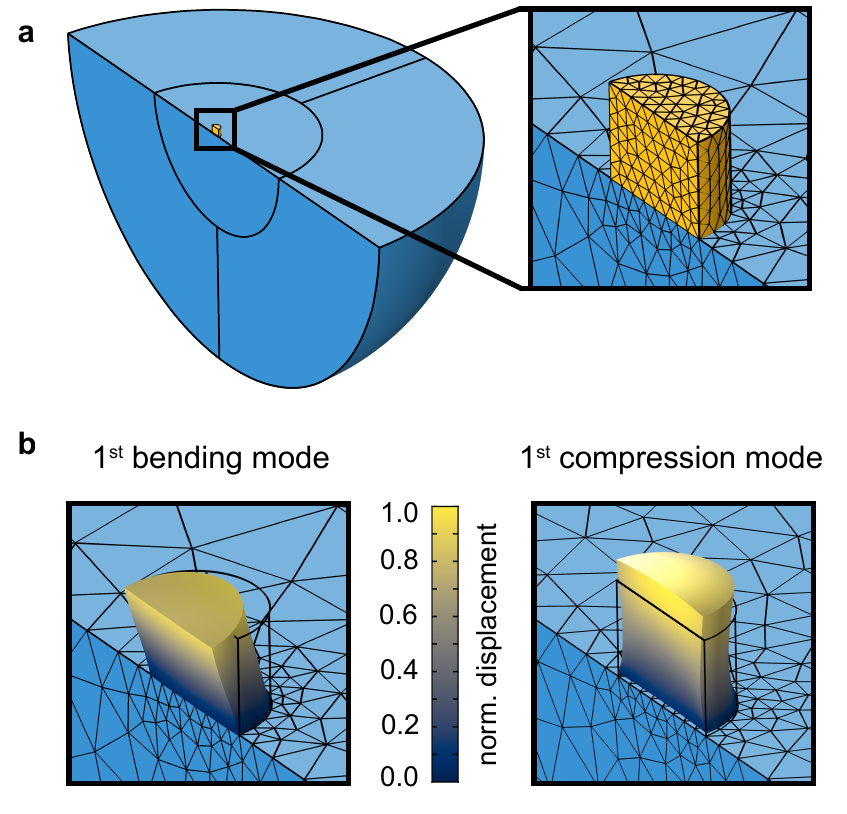}%[width=1\textwidth]
    \caption{Finite element method (FEM) simulations of a single pillar resonator. (a) Geometry and mesh of the FEM simulations. The platinum-carbon pillar is coloured in yellow and the lithium niobate substrate in blue. We reduced the simulated domain to half of the considered domain by exploiting symmetries. (b) Illustration of the shape of the first-order bending and compression mode of the pillar.}
    \label{FEM}
  \end{center}    
\end{figure}

Apart from the pillar, we modeled the substrate as a half-sphere and defined the outer part as perfectly matched layer to mimick an infinitely large substrate. The substrate material was 128$^\circ$ Y-cut LiNbO$_3$. We exploited the symmetry of the lithium niobate crystal\cite{Weis1985} and reduced the simulated domain to half of the considered domain, as shown in Fig.~\ref{FEM}a. Fig.~\ref{FEM}b, c illustrate the shape of the first-order bending and compression mode of the simulated pillar. Both bending and compression modes are actuated by a Rayleigh-type SAW due to its longitudinal and transverse motions\cite{Benchabane2017,Kahler2022}. In our simulations, we focused on the pillar's bending modes because of the optical measurements. The FEM simulations gave an eigenfrequency of \SI{148(54)}{\mega\hertz} for the first-order bending modes and \SI{381(139)}{\mega\hertz} for the second-order bending modes. The specified ranges indicate the minimal and maximal eigenfrequency to be expected based on the uncertainties in the Young's Modulus and the pillar's height. The two measured bending modes of the pillar were around \SI{160}{\mega\hertz} and \SI{167}{\mega\hertz}, which correspond to a Young's modulus of the pillar of around $E = (29\pm5)~\text{GPa}$. The comparison between the simulated and measured eigenfrequencies suggests that we detected the first-order bending modes of the pillar.

We also performed FEM simulations of the thin and the midsize pillar and searched for eigenmodes of both pillars with eigenfrequencies inside the frequency range of the IDTs from around \SI{225}{\mega\hertz} to \SI{335}{\mega\hertz}. We set the  Young's modulus of the pillars to $E =\SI{29\pm5}{\giga\pascal}$, as determined above. We found two eigenmodes for each pillar in the frequency range of the IDTs. The thin pillar vibrates at $f_0 = (283\pm30)~\text{MHz}$ in the first-order compression mode and at $f_0=(351\pm43)~\text{MHz}$ in the third-order bending mode. The midsize pillar vibrates around the same frequency $f_0 = (275\pm28)~\text{MHz}$ as the thin pillar in the first-order compression mode and around $f_0 = (219\pm25)~\text{MHz}$ in the second-order bending mode.

\begin{figure}
  \begin{center}    
    \includegraphics{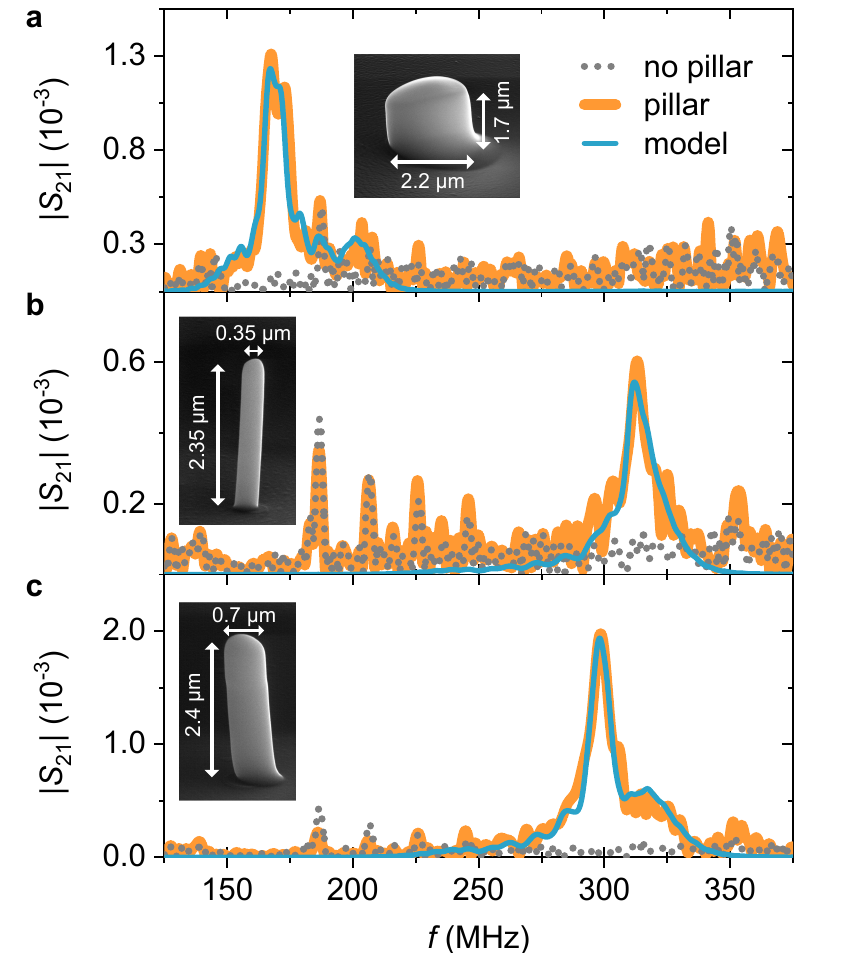}%[width=1\textwidth]
    \caption{Frequency response of pillar resonators transduced by surface acoustic waves (SAW). The pillars scatter an incident SAW towards an interdigital transducer. Around the resonance of the pillars, the scattering of the incident SAW is stronger which result in an increased scattering parameter $S_{21}$. Scanning electron microscope images show the geometry of the measured pillars. The widest pillar, shown in (a), vibrates in the first-order bending mode and the thinner pillars, shown in (b) and (c), in the first-order compression mode. We fitted Eq.~(\ref{S21simple}) to the data with $S_\text{P}$ given by Eq.~(\ref{Sp}).}
    \label{SAWReadout}
  \end{center}    
\end{figure}

\section*{SAW measurements}
In the following, we discuss the frequency responses of the thin, the midsize, and the wide pillar measured by the SAW transduction scheme. We start with the measurements of the wide pillar. In Fig.~\ref{SAWReadout}a two frequency responses are displayed: a measurement of the device with the wide pillar and a measurement  of an identical device without any pillar. Only the device with the pillar shows a peak well above the noise level, which confirms that we measured an eigenmode of the pillar. We analysed the frequency response of the device by fitting $|S_{21}|$, as defined by Eq.~(\ref{S21simple}), to the data. We described the scattering parameter of the pillars $S_\text{P}$ as given by Eq.~(\ref{Sp}) and determined $S_\text{IDT}^2$ by a measurement of a device with two focused IDTs facing each other (see Supporting Data section S2). The IDTs were designed equivalently to the orthogonally arranged IDTs used for the pillar measurements. We decided to fit $|S_\text{21}|$ instead of $|S_\text{21}|/S_\text{IDT}^2$, since the normalization results in a drastically increase of the noise outside the IDTs' frequency range. The result of the fitting is shown in Fig.~\ref{SAWReadout}a and is in excellent agreement with the measured data. The model gives an eigenfrequency of $f_0=169~\text{MHz}$ and a quality factor of $Q=39$. The comparison to the optical measurements discussed above shows that we measured the first-order bending mode of the wide pillar along the y-direction. However, we were not able to detect the bending mode of the pillar along the x-direction. A swap of emitter and receiver IDT gives the same result due to the reciprocity of the device\cite{Morgan2007}. Hence, the bending mode in x-direction does not emit a significant intensity in the direction of the detection IDT and is only weakly actuated by the same. 

In Fig.~\ref{SAWReadout}b, c, the frequency response of the thinner pillars are given in comparison to measurements of identical devices without a pillar. The results clearly show that we measured eigenmodes of the pillars. We fitted Eq.~(\ref{S21simple}) to the data as described above. The model agrees well with the measurements and gives for the thin and midsized pillar an eigenfrequency of around $f_0=311~\text{MHz}$ and $f_0=298~\text{MHz}$, with quality factors of $Q=46$ and $Q=43$, respectively. Both pillars vibrate around the same eigenfrequency despite their difference in diameter by a factor of two. This suggests that we measured the compression mode of both pillars, since the eigenfrequency of compression modes is mainly a function of the height of a pillar and not its diameter\cite{Schmid2016}. These results are in agreement with the FEM simulations discussed above.

\section*{Conclusions}
In conclusion, we demonstrated an electromechanical transduction method for nanomechanical pillar resonators. The technique is reminiscent to darkfield microscopy but probing with SAWs. We showed that the SAW transduction method is able to actuate and detect the motion of pillars with significantly different aspect ratios. One of the pillars vibrates in its first-order bending mode and the other two in their first-order compression mode. Our results illustrate the versatility of the SAW transduction, which originates from its working principle. The SAW transduction is based on resonant scattering of a SAW by a mechanical resonator. As a result, the SAW is not limited by the conductivity of the resonator's material or to a specific resonator geometry. This enables adjustment of the resonator for different sensing purposes. A limit of the SAW transduction is the requirement for a piezoelectric material beneath the IDT's for SAW generation and detection, which can put constrains on the fabrication process. For the propagation of Rayleigh waves no piezoelectricity is required.%A limit of the SAW transduction is the requirement for a piezoelectric material beneath the IDT's for SAW generation and detection, which can put constrains on the fabrication process.

We operated at a maximum frequency of around \SI{300}{\mega\hertz}. However, the SAW transduction scheme is scalable. Commercial SAW devices are usually operated at frequencies of several \SI{10}{\mega\hertz} to several GHz. For example, an increase of the SAW's frequency to \SI{2}{\giga\hertz} would enable the measurement of pillar resonators with diameters of around \SI{50}{\nano\meter} and heights of around \SI{360}{\nano\meter}, exemplifying the potential of the SAW transduction as an access to mechanical resonators on the nanoscale. We expect that the presented SAW transduction method enables specific applications of nanomechanical pillar resonators, e.g., for mass spectrometry or high-speed atomic force microscopy. Moreover, it can facilitate developments and research in other fields, such as quantum information technology and acoustic metamaterials.

%%Bibliography

	%% A small distance to the other stuff in the table of contents (toc)
	%\addtocontents{toc}{\protect\vspace*{\baselineskip}}

%\bibliographystyle{nature}%plain / apalike / amsalpha /

\section*{Authors contribution}
H.K. conceptualized the SAW transduction scheme, performed the electrical measurements, analysed the data, and wrote the original draft. H.A. supervised the electrical measurements, their analysis and supported the analysis of the optical data. R.W. fabricated the pillar resonators under the supervision of H.P.. R.G.W. performed the optical measurements with the support of H.K. and was involved in the analysis of the optical data. I.I. prepared the device for the optical measurements. H.K. wrote the paper with input from all authors. S.S. helped conceptualize and supervised the project. All authors reviewed and edited the manuscript.

\section*{Notes} 
The authors declare no competing interests.

\begin{acknowledgement}
We thank M. Buchholz for the fabrication of the IDT structures and A. Muhamedagić for illustrating the SAW transduction scheme. This work is supported by the European Research Council under the European Unions Horizon 2020 research and innovation program (Grant Agreement-716087-PLASMECS). The financial support by the Austrian Federal Ministry for Digital and Economic Affairs and the National Foundation for Research, Technology and Development is gratefully acknowledged (Christian Doppler Laboratory DEFINE).
\end{acknowledgement}

\newpage

\renewcommand{\figurename}{Supplementary Fig.} %Rename Figures

\section{Supporting Data}

\subsection*{S1: Theory of the optical detection method}
We used the optical detection scheme presented by Molina et al.\cite{Molina2020} to detect the motion of the wide pillar and mapped the amplitude of the optical signal as function of the laser position for two frequencies. The results are presented in the main text. In contrast to Molina et al.\cite{Molina2020}, the diameter of our pillar is wider than the spotsize of the gaussian beam. For this reason, we can not directly compare our measured amplitude maps with the results of Molina et al.\cite{Molina2020}. However, Molina et al.\cite{Molina2020} gave a theoretical model for the generation of the optical signal, which we used for comparison. The result of the applied model can be seen in Supplementary Fig.~\ref{Reflectivity}a-c. Supplementary Fig.~\ref{Reflectivity}a shows the total power of the reflected light when the pillar is not actuated. For this measurement, we used an APD410A/M DC avalanche photodiode from THORLABS.
\begin{figure}
  \begin{center}   \includegraphics{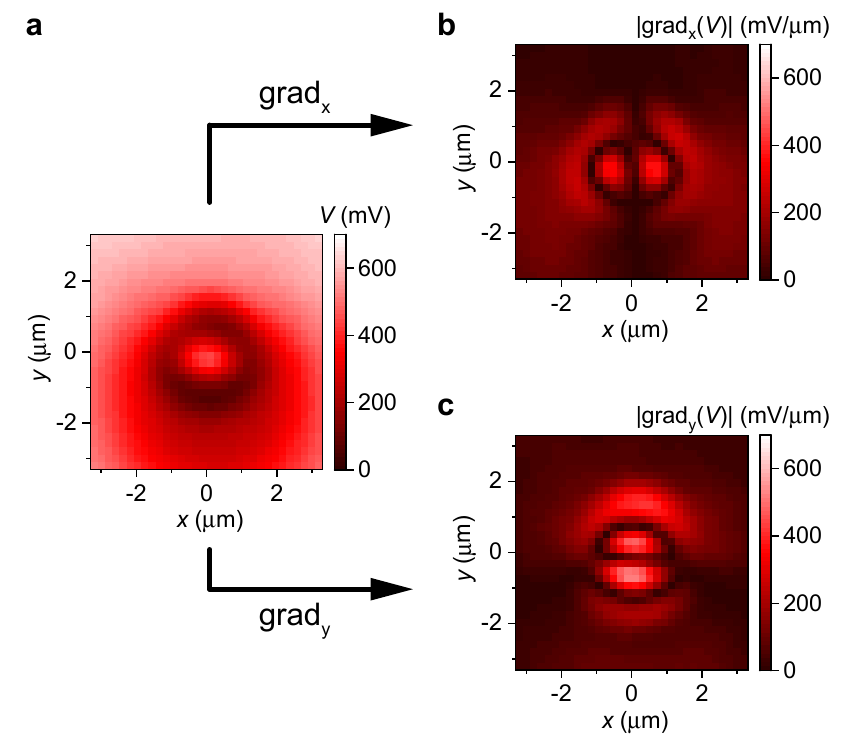}%[width=1\textwidth]
    \caption{Optical detection model. (a) Total power of the reflected light, given by the output voltage $V$ of the photodiode used for detection, as a function of the laser position. The pillar is located around the center of the map and is not actuated. (b), (c) Gradients of the reflected power along the x- and y-axis, which correspond to the vibration axis of the pillar.}
    \label{Reflectivity}
  \end{center}    
\end{figure}
Having in mind that the pillar has a diameter of \SI{2.2}{\micro\meter}, it can be seen that the pillar mainly scatters the incident light at its edge. Based on the model of Molina et al., the maps of the optical amplitude signal are given by the gradient of the total reflected power along the vibration axis of the pillar. The gradients of the reflected power are given in Supplementary Fig.~\ref{Reflectivity}b,c and agree well with the measured amplitudes maps shown in the main text. Hence, the optical signal seems to be mainly generated as predicted by the model of Molina et al.\cite{Molina2020}.

\begin{figure}[t]
  \begin{center}    
    \includegraphics{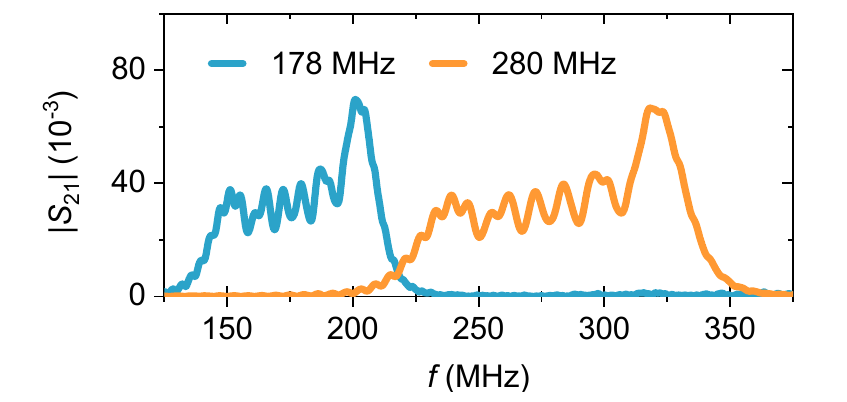}%[width=1\textwidth]
    \caption{Frequency response of two devices with interdigital transducers (IDTs) of different center frequency. The IDTs were designed equivalently to the orthogonally arranged IDTs discussed in the main text, but were facing each other.}
    \label{SIDT}
  \end{center}    
\end{figure}

\subsection*{S2: Transmission scattering parameter of the IDTs}
In the main text, we present an equivalent circuit model for the SAW transduction scheme. The model treats each of the IDTs as a single two port network, which are described by scattering parameters. To determine the transmission scattering parameter of the IDTs in the equivalent circuit model, we measured two IDTs facing each other. The IDTs were placed along the crystallographic X-axis of the lithium niobate substrate and were designed equivalently to the orthogonally arranged IDTs used for the pillar measurements discussed in the main text. 

The transmission scattering parameter $|S_{21}|$ of two devices with facing IDTs are shown in Supplementary Fig.~\ref{SIDT} as a function of frequency: one device with a central frequency of \SI{178}{\mega\hertz} and another with a central frequency of \SI{280}{\mega\hertz}, like the devices discussed in the main text. Since we measured the transmission scattering parameter of the overall device, the transmission scattering parameter of the IDTs $S_\text{IDT}$ is related to $S_{21}$ by 
\begin{equation}
S_\text{IDT}^2(f) = S_{21}(f) \; . 
\end{equation}

\section{Methods}

\subsection{S3: Device fabrication}
 Fig.~1b in the main text shows a SEM image of one of the devices used in this study. The substrate is black lithium niobate (LiNbO$_3$) with a $128^\circ$~Y-cut orientation and a thickness of \SI{350(20)}{\micro\meter}. On top of the substrate, we fabricated interdigital transducers (IDTs) and a pillar resonator. The IDTs were structured by standard UV lithography, deposited by thermal evaporation of Ti($5~\text{nm}$)/Al($150~\text{nm}$)/Au($5~\text{nm}$) and placed so that their mirror plane is either parallel or perpendicular to the crystallographic X-axis of the lithium niobate substrate. The shape of the electrodes of the IDTs are designed to match the wave surface of the SAW that is emitted by the pillar. We calculated the wave surface based on the results of Kovacs et al.\cite{Kovacs1990}.  
 
 The pillar resonator is fabricated by Focused Electron Induced Deposition (FEBID) using a Quanta 3D FEG dual beam microscope from Thermo Fisher Scientific equipped with a standard gas injection system. The FEBID technique is based on gaseous precursor molecules which are locally dissociated on the substrate surface by a focused electron beam. We used MeCpPt$^\text{IV}$Me$_3$ (CAS: 94442-22-5) as precursor molecule, which we heated to 45$^\circ$ for at least 30 min before deposition. The pillars were deposited at a primary beam energy of \SI{5}{\kilo\electronvolt} and a beam current of \SI{92}{\pico\ampere} by a multi-pass, edge-rounding correcting writing pattern towards the gas flux for optimized pillar shape\cite{Winkler2015}. 

\subsection*{S4: SAW transduction} We used a N5247A PNA-X network analyzer from Keysight for the SAW transduction measurements, which were on wafer calibrated. We applied an input power of $-5~\text{dBm}$ and set an IF bandwidth of \SI{10}{\kilo\hertz}. The output signal generated by the scattered SAW was superimposed by electrical crosstalk between the two IDTs. The SAW signal travels with around $4000~\text{m/s}$ and is much slower than the electrical crosstalk signal\cite{Kovacs1990}. This allowed us to perform a time gating to remove the electrical crosstalk from the output signal.

\subsection*{S5: Optical detection} We used a UHF lock-in amplifier from Zurich Instruments to conduct the optical detection measurements. We applied a voltage of \SI{750}{\milli\volt} to the emitter IDT to drive the pillar resonator by SAWs. The optical setup for detection of the pillar's motion is shown in Fig.~2a in the main text. The laser beam was emitted by a TopMode diode laser from Toptica Photonics at a wavelength of \SI{633}{\nano\meter}, incident on the sample with a radiant flux of \SI{76}{\micro\watt}. We focused the laser beam by an objective (x50) with a long working distance on the surface of the substrate resulting in a spot size of the laser beam of around \SI{1.3}{\micro\meter}. The light reflected from the sample was detected by an APD210 avalanche photodidode from MenloSystems with an AC coupled output, which was connected to the input of the lock-in amplifier.

\subsection*{S6: FEM simulations} We performed the FEM simulations in COMSOL Multiphysics (Version 5.5) as described by K\"ahler et al.\cite{Kahler2022} except for the meshing of the inner part of the substrate. We applied a maximum mesh element size of an eight of the SAW's wavelength for the whole inner substrate. The reason for this is the smaller size of our geometry in comparison to the pillar pair simulated by K\"ahler et al.\cite{Kahler2022}.

\bibliography{PaperBibSAWTrans_V5}

%\clearpage

\end{document}